# Sparse and Optimal Acquisition Design for Diffusion MRI and Beyond

Cheng Guan Koay[a], Evren Özarslan[c,d], Kevin M Johnson[a], M. Elizabeth Meyerand[a,b]

[a]Department of Medical Physics
University of Wisconsin School of Medicine and Public Health
Madison, WI 53705

[b]Department of Biomedical Engineering
University of Wisconsin-Madison
Madison, WI 53705

[c]Section on Tissue Biophysics and Biomimetics
National Institutes of Health
Bethesda, MD 20892

[d]Center for Neuroscience and Regenerative Medicine
Uniformed Services University of the Health
Sciences, Bethesda, MD 20814

*Corresponding author:*
*Cheng Guan Koay, PhD*
*Department of Medical Physics*
*University of Wisconsin School of Medicine and Public Health*
*1161 Wisconsin Institutes for Medical Research (WIMR)*
*1111 Highland Avenue*
*Madison, WI 53705*
*E-mail: cgkoay@wisc.edu*

*Short Title:* Sparse and Optimal Acquisition Design for Diffusion MRI

*Abstract*

**PURPOSE:**

**Diffusion Magnetic Resonance Imaging (MRI) in combination with functional MRI promises a whole new vista for scientists to investigate noninvasively the structural and functional connectivity of the human brain—the human connectome, which had heretofore been out of reach. As with other imaging modalities, diffusion MRI data are inherently noisy and its acquisition time-consuming. Further, a faithful representation of the human connectome that can serve as a predictive model requires a robust and accurate data-analytic pipeline. Here, the focus of this paper will be on one of the key segments of this pipeline—in particular, the development of a *sparse* and *optimal acquisition* (SOA) design for diffusion MRI multiple-shell acquisition and beyond.**

**METHODS:**

**We propose a novel optimality criterion for sparse multiple-shell acquisition and quasi-multiple-shell designs in diffusion MRI and a novel and effective semi-stochastic and moderately greedy combinatorial search strategy with simulated annealing to locate the optimum design or configuration. The goal of our optimality criteria is threefold: first, to maximize uniformity of the diffusion measurements in each shell, which is equivalent to maximal incoherence in angular measurements; second, to maximize coverage of the diffusion measurements around each radial line to achieve maximal incoherence in radial measurements for multiple-shell acquisition; and finally, to ensure maximum uniformity of diffusion measurement directions in the limiting case when all the shells are coincidental as in the case of a single-shell acquisition. The approach we take in evaluating the stability of various acquisition designs is based on the condition number and the A-optimal measure of the design matrix.**

**RESULTS:**

**Even though the number of distinct configurations for a given set of diffusion gradient**

directions is very large in general—e.g., in the order of $10^{232}$ for a set of 144 diffusion gradient directions, the proposed search strategy was found to be effective in finding the optimum configuration. It was found that the square design is the most robust (i.e., with stable condition numbers and A-optimal measures under varying experimental conditions) among many other possible designs of the same sample size. Under the same performance evaluation, the square design was found to be more robust than the widely used sampling schemes similar to that of 3D radial MRI and of Diffusion Spectrum Imaging (DSI).

CONCLUSIONS:

A novel optimality criterion for sparse multiple-shell acquisition and quasi-multiple-shell designs in diffusion MRI and an effective search strategy for finding the best configuration have been developed. The results are very promising, interesting and practical for diffusion MRI acquisitions.

## 1. Introduction

Diffusion Magnetic Resonance Imaging (MRI) [1-3], a noninvasive and non-ionizing magnetic resonance imaging technique, is one of the most exciting and promising imaging techniques and is increasingly being used in basic neuroscience research and clinical studies of neurological diseases [4-8]. Diffusion MRI in combination with functional MRI [9] promises a whole new vista for scientists to explore the human brain from structural and functional connectivity to human cognition. Such an exploration in human inquiry had heretofore been out of reach.

Diffusion MRI data are inherently noisy and its acquisition time-consuming. Hence, a faithful representation of the human connectome [10, 11] that can serve as a predictive model for scientists to make inferences and discoveries requires a robust and accurate data-analytic pipeline [12-17] from data acquisition, data processing to data visualization and data reduction. Here, our focus will be on the sparse and optimal acquisition (SOA) design for diffusion MRI, especially in the context of multiple-shell acquisition.

Diffusion tensor imaging (DTI)[3] is the most commonly used variant of diffusion MRI techniques [18-25] in clinical studies, and consequently, the acquisition design of DTI has also been investigated extensively [26-31]. However, these investigations [26-31] focused mainly on a very special acquisition, which is known as a single-shell acquisition in which the diffusion gradient directions can be conceptualized as vectors that are antipodally symmetric and are also distributed uniformly on some spherical shell because a single diffusion weighting (also known as b-value) is employed in the acquisition (without counting the non-diffusion weighted measurements). With the emerging interests in Compressive Sensing (CS) methodology in imaging sciences [32-34], the question and problem of constructing a SOA design will surely be of significant interests because CS unified the notions of sparsity in signal representation and of incoherence in sampling in a novel attempt to unravel the problem of finding the sparsest signal under the condition of underdetermined set of measurements. We should note that our goal here to construct a SOA design that is sparse and maximally incoherent in sampling so as to achieve low coherent with respect to the typical bases used in 3D reconstruction of diffusion propagator, see e.g., Refs.[35-37]. Maximum incoherence in sampling is a desirable sampling strategy and is related to uniformity of the diffusion gradient vectors (or measurements) in angular sampling for a single-shell acquisition.

Even though we briefly touched on the issue of and a simple solution to optimal ordering for multiple-shell acquisitions in our recent work [31], we left the question regarding the optimal placements of diffusion gradient directions across and within spherical shells unaddressed because of the inherent complexity of the problem of maximizing the incoherence for a given number of measurements on these spherical shells, which is a combinatorial optimization of substantial complexity due to enormous number of distinct configurations, e.g., $10^{27}$ to more than $10^{232}$ for diffusion MRI and $10^{46642}$ for three-dimensional radial MRI, see Appendix A. The focus of this work is to provide a novel and practical solution to the problem of SOA design in diffusion MRI.

Multiple-shell acquisition, which is not uncommon in DTI studies, is increasingly being used in other diffusion MRI techniques [36, 38] but the search for its optimality criterion has been elusive. In this work, we propose a novel optimality criterion for sparse multiple-shell and quasi-multiple-shell acquisition designs in diffusion MRI and an effective semi-stochastic and moderately greedy combinatorial search strategy with simulated annealing [39] to locate the optimum design or configuration.

Our proposed method determines the optimal sampling pattern for multiple-shell acquisition or quasi-multiple-shell acquisitions through optimization of a cost function once the required parameters, specifically the number of shells (or b-values [3] or q-values [8, 40]) and the number of measurements in each shell, are provided. The cost function directly incorporates the criteria to (1) maximize uniformity of the diffusion measurements in each shell, which is equivalent to maximal incoherence in angular measurements, (2) maximize coverage of the diffusion measurements around each radial line to achieve maximal incoherence in radial measurements for quasi-multiple-shell acquisition, and finally (3) ensure maximum uniformity of diffusion measurement directions in the limiting case when all the shells are coincidental as in the case of a single-shell acquisition. The results are very exciting and promising and they showed that the search strategy is very effective in finding the optimum configuration. Further, it was found that the square design is the most robust (i.e., with stable condition numbers and A-optimal measure across varying experimental conditions) among many other possible designs of the same sample size. Note that the lower the A-optimal measure the lower the mean squared error of the model [41].

## 2. THEORY

Suppose that we have the required parameters such as the number of shells, denoted by $K$, and the number of measurements in each shell, i.e., $j$-th shell has $n_j$ number of measurements so that the total number of measurement, denoted by $N$, is given by $N = \sum_{i=1}^{K} n_i$.

To construct a cost function that directly incorporates the following criteria:

(1) maximize uniformity of the diffusion measurements in each shell;

(2) maximize coverage of the diffusion measurements around each radial line;

(3) ensure maximum uniformity of diffusion measurement directions in the limiting case when all the shells are coincidental as in the case of a single-shell acquisition, we will first deal with criterion #3 by having all $N$ measurements (or antipodally symmetric points) uniformly distributed on the upper hemisphere of unit radius. This strategy ensures sufficient angular coverage in the case of multiple-shell or quasi-multiple-shell acquisitions and maximum angular coverage in the limit when the shells are coincidental as in the case of a single-shell acquisition. Note that, criteria #2 and #3 would be incompatible if we were to maximize coverage of the diffusion measurements *along* rather than *around* each radial line. We should point out that our operational meaning of a quasi-multiple-shell acquisition is similar to multiple-shell acquisition as described above except that the points in each shell may be associated with slightly different diffusion weightings from the nominal weighting for that shell. Suppose we have three shells, with nominal b-values of 500 s/mm$^2$, 1000 s/mm$^2$, and 1500 s/mm$^2$, and each shell has 20 uniformly distributed measurements, a quasi-multiple-shell acquisition would be an acquisition in which the 20 measurements of the first, the second and the last shells have b-values (in units of s/mm$^2$) ranges from 250 to 750, 750 to 1250 and 1250 to 1750, respectively. Such an approach in designing radial measurement will ensure incoherent sampling radially while maintaining maximal angular incoherence. Next, we will develop a cost function that will identify specific points to be moved to various shells so as to satisfy criteria #1 and #2 simultaneously.

To facilitate subsequent discussion, we will introduce a convenient graphical representation as shown in Figure 1 with the assumptions that $n_1 \geq n_2 \geq \cdots \geq n_K$ and that the first row has $m_1$ measurements, the second row has $m_2$ measurements and so on such that $N = \sum_{i=1}^{n_1} m_i$ and $K = m_1 \geq m_2 \geq \cdots \geq m_{n_1}$. The computation of $m$'s can be easily achieved through algorithmic approach by putting ones in the jagged grid and summing the values across the columns for each row. The points placed on the jagged grid may be taken to be any permutation from a list of $N$ original points, $\{P_1, \cdots, P_N\}$, and the permutated list of points are denoted by $P_{S_k}$ with a single index, $S_k$, with $k$ ranges from $1$ to $N$. Hence, the sequence, $\{S_1, S_2, S_3, \cdots, S_N\}$, is a permutation of the original sequence, $\{1, 2, 3, \cdots, N\}$. Each point, $P_{S_k}$, is a unit vector. For convenience, we also use the same symbol but with two indices, i.e., $P_{ij}$, to refer to the point located at the $i$-th row and the $j$-th column in Figure 1; here, the assignment, denoted by the symbol $\leftarrow$, is made from $P_{S_k}$ to $P_{ij}$ in a column-wise fashion, i.e., $P_{11} \leftarrow P_{S_1}$, ..., $P_{n_1 1} \leftarrow P_{S_{n_1}}$, $P_{12} \leftarrow P_{S_{n_1+1}}$ and so on with the final assignment given by $P_{n_K K} \leftarrow P_{S_N}$.

Due to the constraint and the important role of antipodal symmetry in diffusion MRI, we will also introduce a simplifying strategy by classifying points as real and their corresponding antipodal points as virtual. If we have $N$ real points (on the upper hemisphere), denoted by unit vectors $r_i$ with $i = 1, \cdots, N$, then the total electrostatic energy for the complete configuration of $2N$ points of both real and virtual points is given by:

$$\varphi = \frac{N}{2} + 2 \sum_{i=1}^{N-1} \sum_{j=i+1}^{N} \left( \frac{1}{r_{ij}} + \frac{1}{\sqrt{4 - r_{ij}^2}} \right) \tag{1}$$

with $r_{ij} \equiv \| r_i - r_j \|$. Note that Eq.[1] is expressed solely in terms of real points. If we define $S(r_i, r_j) \equiv \frac{1}{r_{ij}} + \frac{1}{\sqrt{4 - r_{ij}^2}}$ then $S(r_i, r_j)$ may be thought of as a reciprocal metric (or reciprocal distance measure) between two real points. Now, we are ready to introduce the cost function, which can be express concisely as:

$$\Phi = \cfrac{\displaystyle\sum_{j=1}^{K} \underbrace{\cfrac{\overbrace{\dfrac{n_j}{2} + 2\sum_{h=1}^{n_j-1}\sum_{i=h+1}^{n_j} S(P_{hj}, P_{ij})}^{\phi_R(j)} - E_{n_j}}{E_{n_j}}}_{\tilde{\phi}_R(j)}}{\displaystyle\sum_{q=1}^{n_1} \underbrace{\cfrac{\overbrace{\dfrac{m_q}{2} + 2\sum_{r=1}^{m_q-1}\sum_{s=r+1}^{m_q} S(P_{qr}, P_{qs})}^{\phi_C(q)} - E_{m_q}}{E_{m_q}}}_{\tilde{\phi}_C(q)}} \qquad [2]$$

The function $\phi_R(j)$ denotes the electrostatic energy of all the points in the $j$-th column. Similarly, $\phi_C(q)$ denotes the electrostatic energy of all the points in the $q$-th row. Note that $E_n$ is the electrostatic energy of the most uniformly distributed antipodally symmetric point set with $n$ number of real points. Therefore, $\tilde{\phi}_R(j)$ (or $\tilde{\phi}_C(q)$) is the relative difference between $\phi_R(j)$ and $E_{n_j}$ (or between $\phi_C(q)$ and $E_{m_q}$). Minimizing $\Phi$ is equivalent to simultaneously minimizing the numerator of Eq.[2] and maximizing the denominator of Eq.[2]. Minimizing the numerator is aimed at fulfilling criterion #1 while maximizing the denominator is aimed at fulfilling criterion #2. The motivation behind the use of relative difference in evaluating the electrostatic energy is to facilitate comparison among configurations of different sizes. For example, it allows us to determine whether a 5-point set is "more uniform" than a 7-point set in a quantitative and comparable manner.

Minimizing $\Phi$ is a challenging combinatorial optimization problem that requires rearrangement of points to achieve lower $\Phi$ value. It can be shown that the number of distinct configurations is $N!$, $N! \equiv n \cdot (n-1) \cdots 2 \cdot 1$ is the factorial function, if the numbers of measurements in each row and each column are distinct. For the special case when the grid is rectangular with $n$ rows and $m$ columns (shells), the number of configuration is given by $\dfrac{(m \times n)!}{m! n!}$. The number of distinct configurations is astronomical for typical values of $m$ and $n$, which renders exhaustive search impractical. For example, with $m = 8$ and $n = 30$, i.e. $N = 240$, which is typical in diffusion MRI applications, the number of distinct configurations is in the order of $10^{431}$. Therefore, an effective and efficient search strategy capable of finding the optimal configuration within appropriate timeframe is imperative. We will present a semi-stochastic and moderately greedy algorithm with simulated annealing in the next section to solve this interesting but challenging combinatorial problem whose solution will shed new light on many areas of research beyond diffusion MRI such as geosciences, imaging sciences and notably three-dimensional radial MRI, please refer to Appendix A in which we showed how the proposed cost function and search strategy can be adapted to solve the problem of optimal view-ordering in three-dimensional radial MRI.

## 2.2 Semi-stochastic and moderately greedy combinatorial search strategy

We propose a novel search strategy to seek the optimal configuration. The algorithm is outlined below:

**Semi-stochastic and moderately greedy combinatorial search algorithm**

**Preliminary step:** Given an ordered list of $N$ number of points, $\{P_1, \cdots, P_N\}$, obtained from tabulated data or from Refs.[31, 42], we make a one-to-one correspondence between $\{P_1, \cdots, P_N\}$ and $\{1, 2, 3, \cdots, N\}$. It is assumed that we also have the relevant information such as the number of shells, $K$, and the number of measurements in each row and column, i.e., $n_1 \geq n_2 \geq \cdots \geq n_K$ and $K = m_1 \geq m_2 \geq \cdots \geq m_{n_1}$. We will find the maximum and minimum values of $\Phi$ from 5000 random configurations. We then set $T$, which is the effective temperature to be used in our modified simulated annealing, to be 5 times the difference between these maximum and minimum values. The first $T$ will be denoted by $T_0$.

Fill the jagged grid with a random permutation of $\{1, 2, 3, \cdots, N\}$ column by column,

**Step 1:** Find the index $j*$ for which $\tilde{\phi}_R(j)$ is maximum.

**Step 2:** Find the index $q*$ for which $\tilde{\phi}_C(q)$ is minimum.

**Step 3:** If $P_{q*j*}$ exists, we look for lower energy configuration by swapping it with other points systematically in the jagged grid. If the rearranged configuration leads to lower $\Phi$, exit and set the "flag" to 1. If none of the rearrangements leads to lower $\Phi$, then exit and set the "flag" to -1.

If $P_{q*j*}$ does not exist, we suggest using a point within $q*$-th row and a point within $j*$-th column and swap these points with other points in the grid. Specifically, we look for the point, $P_{q*r}$, that *minimizes* the following inner sum:

$$\sum_{\substack{s=1 \\ s \neq r}}^{m_{q*}} S(P_{q*r}, P_{q*s}),$$

which is equivalent to finding a point, $P_{q*r}$, that is furthest away from those in the $q*$-th row .Similarly, we may also look for a point, $P_{ij*}$ that *maximizes* the following inner sum:

$$\sum_{\substack{k=1 \\ k \neq i}}^{n_{j*}} S(P_{ij*}, P_{kj*}),$$

which is equivalent to finding a point, $P_{ij*}$, that is closest to those in the $j*$-th column. Then, we can successively swap these two points, $P_{q*r}$ and $P_{ij*}$, with all other points in the grid to search for a better configuration. If the rearranged configuration leads to lower $\Phi$, exit and set the "flag" to 1. If none of the rearrangements leads to lower $\Phi$, then exit and set the "flag" to -1.

**Note: Step 4a and Step4b below are alternative strategies considered in this paper but Step 4b is the strategy of choice for its effectiveness in finding optimum configuration.**

**Step 4a:** If the flag is -1, two distinct points on the grid will be randomly selected and swapped (or two pairs of points may be randomly selected and swapped), if the rearranged configuration is better than the current configuration then we go back to Step 1, otherwise we will perform the same operation of (internal) randomization for another 500 or more times before exiting with a flag of -1. We should note that internal randomization degrades efficiency but improves the overall quality of the search.

**Step 4b:** If the flag is -1, we adopt the simulated annealing approach in which configurations with higher energy (or cost function value) may still have the likelihood of being accepted if the effective temperature is high.

First, two distinct points on the grid will be randomly selected and swapped, if the rearranged configuration is better than the current configuration then we set the flag to "1" and go to Step 5, otherwise we accept the new configuration with probability

$$\exp(-(\Phi_{new} - \Phi_{old})/T),$$

where $\Phi_{new}$ is the value of the cost function of the new configuration and $\Phi_{old}$ is that of the old configuration.

Internally, we keep a counter to ensure that no more than $100N$ trials for each value of the effective temperature in which the new configuration is accepted. If the value of the counter reaches $100N$, we reset the counter to zero and reduce the effective temperature according the simple formula:

$$T_{n+1} = 0.9T_n,$$

where $T_n$ is the current effective temperature and $T_{n+1}$ is the next effective temperature to be used in simulated annealing.

Note that Step 4b will be repeated until either we find a new $\Phi_{new}$ that is truly lower than $\Phi_{old}$ and at which point we set the flag to be "1" and move on to Step 5 or the number of repetitions of Step 4b reaches a predefined number, which is the maximum iteration allowed (e.g., we used 1000000), and the flag will be set to "-1".

**Step 5:** If the "flag" is 1 then repeat Steps 1, 2, 3 and 4. Otherwise, exit the algorithm.

We should note that Step 4a and Step 4b are two alternative strategies and the search strategy that incorporates Step 4a is slightly greedier than that of the search strategy that incorporates Step 4b. Based on our numerical testing, we found that the search algorithm with Step 4b, which is based on simulated annealing, is more effective in finding the optimum solution.

Finally, we will touch on some steps that could significantly speed up the search. Since the distances between the original spherical points are fixed, we should compute and store the values of all the inverse distances, i.e., $S(P_i, P_j)$, in a form of a lower triangular matrix, see Figure 2. It should also be clear by now that it is more efficient to consider rearrangement of the sequence, $\{S_1, S_2, S_3, \cdots, S_N\}$, in, $P_{S_k}$, than to actually rearrange or move the points on the grid. Therefore, the final product from the search algorithm is the optimum configuration, which is simply a grid with numbers rearranged.

## 2.3 Optimal Ordering Strategy

The key motivation behind the use of optimal ordering in diffusion MRI measurements (or gradient directions) is due in part to the desire of salvaging as much information as possible from partially completed diffusion MRI scan in the event that the scan has to be interrupted unexpectedly, e.g., see Refs.[29, 30]. Our recently proposed deterministic optimal ordering strategy for the single-shell acquisition [31] is immediately applicable to the proposed multiple-shell acquisition because of criterion #3. Here, we briefly outline the strategy of incorporating our previously proposed optimal ordering in this work. Since the points are originally placed uniformly on the same sphere, our recent deterministic optimal ordering method can be applied to the point set to create a list of optimally ordered point set, denoted by List #1, which is then feed into the semi-stochastic and moderately greedy combinatorial search algorithm proposed above. The final optimum multiple-shell acquisition, denoted by List #2, can be written out as a text file with points and their corresponding b-values or q-values listed side by side. The points in List #2 is arranged in exactly the same order as in List #1 but their corresponding b-values retrieved from the optimum configuration (or the grid, the output from the above search algorithm).

### 3. Method of Evaluation of Acquisition Designs

The approach we take in evaluating the stability of various acquisition designs is based on the condition number and the A-optimal measure of the design matrix. The use of condition number in the analysis of stability of linear inverse problem is well known, see Refs. [28, 43, 44]. To construct the design matrix, we adopt the basis functions proposed in Ref. [37]. There are several proposed reconstruction techniques [35-37] but the basis functions proposed in Ref. [37] is better suited for the analysis diffusion MR signals. We have made a few modification to the basis functions so that the basis functions are real-valued and orthonormal.

In brief, the three-dimensional normalized q-space signal can be expressed in terms of the modified version of the orthonormal basis wave functions used in three-dimensional harmonic oscillator in quantum mechanics [37, 45] as follows:

$$E(\mathbf{q}) = \sum_{N=0}^{N_{\max}} \sum_{\substack{\{l+2j=N+2\} \\ j \geq 1, \ l \geq 0}} \sum_{m=-l}^{l} a_{jlm} \Psi_{jlm}(u, \mathbf{q}) \qquad [3]$$

with

$$\Psi_{jlm}(u, \mathbf{q}) = R_{jl}(u, \mathbf{q}) \, Y_{lm}(\theta, \phi),$$

$$R_{jl}(u, \mathbf{q}) = \sqrt{2}(2\pi u)^{l+\frac{3}{2}} \sqrt{\frac{(j-1)!}{\Gamma(l+j+1/2)}} q^l \exp(-\frac{(2\pi u)^2}{2} q^2) L_{j-1}^{l+1/2}((2\pi u)^2 q^2),$$

$$Y_{lm}(\theta, \phi) = \begin{cases} -\sqrt{\dfrac{2l+1}{2\pi} \dfrac{(l+m)!}{(l-m)!}} \sin(m\phi) P_l^{|m|}(\cos(\theta)), & -l \leq m \leq -1 \\ \sqrt{\dfrac{2l+1}{4\pi}} P_l(\cos(\theta)), & m = 0 \\ \sqrt{\dfrac{2l+1}{2\pi} \dfrac{(l-m)!}{(l+m)!}} \cos(m\phi) P_l^m(\cos(\theta)), & 1 \leq m \leq l \end{cases}$$

Note that $\mathbf{q} = q\hat{\mathbf{q}}$, $\hat{\mathbf{q}} = [\sin(\theta)\cos(\phi), \sin(\theta)\sin(\phi), \cos(\theta)]^T$, $L_n^l$ is the generalized Laguerre polynomial of order n, $P_l^m$ is the associated Legendre polynomial and $\Gamma$ is the Gamma function. Note also that the second summation is supposed to be evaluated for all even number $l \geq 0$ and integer $j \geq 1$ such that $l + 2j = N + 2$. Further, the definition of the real-valued spherical harmonic functions, $Y_{lm}$, is consistent with our previous work, see [16]. Given a collection of q-vectors (measurements), it is clear that Eq.[3] with evaluation at different q-vectors can be formulated as a matrix equation of the following form,

$$\mathbf{y} = \mathbf{\Phi}\mathbf{\beta},$$

where the elements of the observation vector, $\mathbf{y}$, are the q-space signal, $E(\mathbf{q})$, measured at different q-vectors, and the elements of $\boldsymbol{\beta}$ are the coefficients $a_{jlm}$ in a particular order; this particular order also affects how the basis functions are ordered in the design matrix, $\boldsymbol{\Phi}$. Finally, we would like to remind the reader that the condition number of $\boldsymbol{\Phi}$ is the ratio of the largest to the smallest singular values while the A-optimal measure is the matrix trace of the inverse of the moment matrix, i.e., $tr((\boldsymbol{\Phi}^T\boldsymbol{\Phi})^{-1})$.

**4. RESULTS**

**4.1 Illustrative Example**

We implemented the above cost function and the proposed search strategy in Java on a machine with an Intel® Core™ i7 CPU at 1.73 GHz. We should note that the best electrostatic energy values from a single-shell configuration with different sample sizes will be needed to test our proposed approach. However, we have made available these values on the web. The first 365 configurations were iteratively optimized by using the deterministically generated antipodally symmetric point set [42] as the initial solution. The resultant electrostatic energy values have subsequently been tabulated and are available through the web, see the Acknowledgment for the URL. Electrostatic values for sample size beyond 365 are currently computed approximately from the deterministic point set. Eventually, more values beyond the first 365 will be based on point set derived from iteratively optimization.

We would like to introduce a simple and practical example to illustrate the feasibility of our approach and some important features in the results that are universal across different sample sizes. In this example, we begin with a simple 12 by 12 grid and generated 50000 independent trials by filling up the grid with any random permutation drawn from (1,2,...,144); each number refers to the location of a point in a list of 144 iteratively optimized points generated, tabulated and stored as a text file. We should mention that the number of distinct configurations is astronomically large and it is in the order of $10^{232}$; this number is greater than the number of atoms in the observable universe, which is in the order of $10^{80}$. To depend on exhaustive search to find the optimum configuration is to wait for eternity; it would still take us $10^{201}$ millennia if we could exhaustively search $10^{20}$ configurations in one second.

The distribution of the electrostatic energy of these random configurations is shown as the red histogram in Figure 3. The proposed method (using Step 4a and with 500 internal randomization) was applied to these configurations in an attempt to minimize the electrostatic energy. The blue histogram is the resultant distribution. Please refers to the inset to contrast and compare these two different histograms more closely. The lowest cost function value was 0.00369.

Separately and independently, we performed only one single trial using the proposed method with Step 4b (i.e., simulated annealing) and found that the lowest cost function value to be 0.00297. However, this single trial took approximately the same amount of time in execution, which was about 222 minutes 53 seconds, as compared to the 50000 trials using the other alternative approach (i.e. with Step 4a). Even though, the computational cost is higher, we believe the proposed method with the incorporation of simulated annealing approach in Step 4 is more effective in finding a more optimal configuration.

In what follows, we will present the point set from the most optimum configuration in two distinct perspectives. First, the points in the same column are color-coded similarly and points in different columns have distinct hues, see Figure 4. It is clear that there is no sign of any cluster of points with the same hue. We should remind the reader that the points of the same color are on the same spherical shell and therefore should be as far apart from each other as possible to achieve maximal angular incoherence. Finally, the points in the same row are color-coded similarly and points in different rows have distinct hues, see Figure 5. It is clear that we should expect to see 12 clusters of points with points in each cluster having the same hue. Again, we remind the reader that the points of the same color are, in this case, to be moved to distinct spherical sphere to achieve sufficient radial coverage or radial incoherence. This second perspective is the most stringent test of the efficacy of the cost function and of the search strategy because any discrepancy in the cluster can be easily detected by visual inspection.

Another way of inspection that is more convenient for large sample size is to investigate the variability of the values of $\phi_R(j)$ for all the columns and $\phi_C(q)$ for all the rows when the grid is rectangular, or of $\tilde{\phi}_R(j)$ and $\tilde{\phi}_C(q)$ for jagged grid; we used this approach in another example in Appendix A.

**4.2 Evaluation of Acquisition Designs**

For simplicity, the number of measurements was chosen to 81 to ensure that a square grid of 9x9 was one of the 18 designs considered in this investigation, see Table 1. We should note that the number of possible designs is equal to the unrestricted partitions of the integer 81, which turns out to be 18004327, see Ref. [46].

To construct the design matrix and evaluate its condition number for different experimental conditions, we have to have the following information:

1. A collection of measurement vectors, $\mathbf{q}$'s. The q-values are shown in Table 1. In each design shown in Table 1, the orientation information of the measurement vectors was obtained by the method proposed in Section 2.

2. The value of $u$ in Eq.[3]. The inherent advantage of the basis function proposed by Özarslan [37] is that $u$ can be associated with the Einstein relation, i.e., $u = \sqrt{2D\Delta}$, where $D$ is the diffusion coefficient and $\Delta$ is the diffusion time. Due heterogeneity of diffusivity in the human brain, it is important to investigate the stability of the acquisition design, i.e., the condition number of the design matrix under two very distinct values of diffusivity. Here, the two different values of the diffusion coefficient were chosen, the first one is close to the observed value of free diffusion in the human brain, which is $2.1 \times 10^{-3}$ mm$^2$/s and the other is close to very slow diffusion ($2.1 \times 10^{-5}$ mm$^2$/s) as in the case of hindered diffusion.

3. The value of $N_{max}$ was chosen to be 4, which yields exactly 22 coefficients in the series expansion in Eq.[3].

With the relevant parameters set to the chosen values, we evaluated the condition number and the A-optimal measure associated with each design shown in Table 1 (fast diffusion) and Table 2 (slow diffusion). It is clear from the results that any design closer to the square one has a lower combined A-optimal measure and Design #18 (the square design) is has the lowest combined A-optimal measure. The results in Tables 1 and 2 show that the condition numbers of Design #17 and #18 are less variable than other designs. We also investigated the behavior of the condition number of the design matrix of Design #18 as a function of the diffusion time with diffusivity fixed at free diffusion, see Figure 6. It is interesting to note that there are two minima in Figure 6 but the minimum indicated at 16.32 ms is the one that is more practical for diffusion MRI since the diffusion time has to be much greater than the width of the diffusion gradient pulse. We would like to point out that scaling the diffusion coefficient up or down will increase the condition number of the design matrix if the diffusion time is fixed at 16.32 ms. Finally, we investigated the condition number of two design matrices derived from the radial scheme, i.e., nine uniformly distributed q-vectors on each shell and the q-values used were obtained from Table 1, and the DSI-like sampling scheme [18]. The condition numbers for the radial scheme were $2.5 \times 10^{18}$ and $5.9 \times 10^{16}$ for both cases of fast diffusion and slow diffusion, respectively. The values of the A-optimal measure for the radial scheme were $2.1 \times 10^{23}$ and $1.4 \times 10^{24}$ for both cases of fast diffusion and slow diffusion, respectively. The condition numbers for the DSI scheme were $3 \times 10^{15}$ for the case of fast diffusion and 69.35 for the case of slow diffusion. The values of the A-optimal measure for the DSI scheme were $7 \times 10^{30}$ for the case of fast diffusion and $7 \times 10^{9}$ for the case of slow diffusion. Both of these commonly used sampling schemes, DSI and radial, have a much higher combined value of A-optimal measure than many of the designs considered in Table 1. The DSI sampling scheme can be visualized as integral lattice points that are within and on some radius of integral value, which is associated with the maximum q-value. We obtained 76 lattice points with integral radius of 3 and added 5 more repeated measurements at the center of q-space to ensure the total number of sampling points is exactly the same as other designs considered in this work.

Finally, we note that the time taken by the proposed search strategy to find the square design (Design #18) was about 18 minutes.

## 5. DISCUSSION

We have been grappling with the problem of the SOA design for a multiple-shell acquisition in diffusion MRI for awhile without any progress until we treaded to other territory to address the question on the optimal view-ordering in three-dimensional radial MRI. Chronologically, the solution to the three-dimensional radial MRI was found first and that of diffusion MRI later. Due to our continued interest and research effort in diffusion MRI, we have decided to present our finding in the context of diffusion MRI and made only a brief mention, in a form of an appendix, on how our proposed approach can be adapted to solving optimal view-ordering in three-dimensional radial MRI. However, we should point out that the radial scheme is less optimal than the proposed design, notably the square design, Design #18 in Table 1.

The main objective of this work is to show the simplicity and intuitiveness of our optimality criteria for sparse multiple-shell and quasi-multiple-shell acquisitions in diffusion MRI and to share with the reader a novel search algorithm, which is based on the notion of *moderate greediness*, a term borrowed from social science. Incorporation of simulated annealing in Step 4, specifically Step 4b, of the proposed search algorithm makes the search more effective but at a higher computational cost. Simulated annealing, with its distinctive feature of being able to explore larger number of slightly less optimal configurations (with respect to a particular level of effective temperature) at higher effective temperature, and fewer and fewer of those suboptimal configurations as the effective temperature decreases, may be thought of as *measured sacrifice* to achieve coherence in socio-economic terms. Note that the apparent coherence here in angularity will disappear once the points are moved to different shells so as to achieve better radial incoherence. Even better radial incoherence can be achieved with quasi-multiple-shell design. We believe angular and radial incoherence in sampling will play an important role in diffusion MRI as the number of compressive sensing applications to diffusion MRI increases [47].

The results are very promising, interesting and practical for diffusion MRI acquisitions. It is interesting to note that the square design turns out to be better in terms of lower combined A-optimal measures than all the designs considered in this study and much better than the commonly used radial design, which is similar to the sampling scheme of three-dimensional radial MRI, and the DSI scheme. The key findings are that (1) the multiple-shell design similar to Design #18 is better than the current paradigm of sampling on a single shell, and (2) this particular multiple-shell design is also better than the multiple-shell design obtained from the radial scheme. Beyond the confine of diffusion MRI, it is very interesting to point out the following fact, which is that we managed to find an optimum configuration for a realistic example *within a realistic timeframe* even though the number of distinct configurations is astronomically large. This particular finding is most encouraging and exciting from the point of view of computational science and engineering.

      Finally, we would like to mention that the proposed approach may be of relevance to other applications such as geosciences (network design of underground seismometers), geospatial intelligence (the design of multiple satellite constellations [48]), and last but not least, other biomedical imaging modalities.

**APPENDIX A : Application of the proposed approach to finding optimal view-ordering in three-dimensional radial MRI**

3D MR data is most often acquired with multiple readout windows. The inconsistency among readout windows is usually due to subject motion and variation in spin states. This inconsistency leads to image artifact in the final reconstructed 3D volume. The nature and the magnitude or severity of the artifact are determined by the k-space trajectory and the order in which the views are acquired. The effects of view ordering have been extensively studied in 2D and 3D Cartesian MRI and 2D radial MRI [49-54]. Most existing techniques, aim to isolate signal variations for a single source (e.g. T2 decay) or along a single time dimension.

Unfortunately, the design of view ordering schemes in multiple-dimensions and/or for 3D trajectories is non-trivial. Even though 3D radial MRI [55] is increasingly common in clinical and research studies such as MR angiography, sodium MRI and musculoskeletal MRI [53, 56-60], the lack of an intuitive optimality criterion for designing view ordering has led to the use of various heuristic ordering schemes such as the simple interleaving method, the pseudo-random bit-reversal method [61], and the golden means method [62]. While these schemes have been shown to produce images of good quality, there is no guarantee that they are optimal.

In this appendix, we will present a novel view-ordering optimality criterion for 3D radial MRI that builds upon the proposed cost function and search strategy in the main text for finding optimal view-ordering. Our proposed method determines view ordering through optimization of a cost function. The cost function directly incorporates criteria to minimize coherent aliasing from missing or corrupt data. The proposed method can be illustrated with an simple example drawn from 3D radial fast spin echo (FSE) imaging technique (or rapid acquisition with relaxation enhancement [RARE])[63].

FSE has led to significant reduction in scan time with recent developments on variable refocusing flip angles [64, 65], which have made it possible to acquire MR data with very long echo trains. The reduction in scan time may also be used to improve image resolution. However, the acquisition of multiple echoes within the same excitation comes at a cost of enhanced image artifacts such as blurring or ringing because of T2 decay. T2 decay introduces signal modulation in k-space. Therefore, different view-ordering strategies [49-54] have been developed and used in an effort to make the signal modulation as incoherent as possible in k-space. Our view-ordering strategy is consisted of two steps. The goal of the first step is to make the signal modulation as incoherent as possible in each echo train and among the same echoes across the trains. The goal of the second step is to make the signal modulation as incoherent as possible in different echo trains and in different echoes across the trains.

Suppose that we have a set of $N$ points uniformly distributed on the surface of the unit sphere, e.g., Ref. [66], and $N$ is a composite number, e.g., $N = m \times n$. Here, we desire to sample $N$ points from the set of points $P$, and only need to know the order in which to sample them, $S$. The view ordering, $S$, can be separated into multiple dimensions. For example, $S$ may be split into $n$ echo trains, each $m$ echoes long. We can imagine placing points on a $m \times n$ grid with $m$ rows and $n$ columns, which is a special case of our general jagged grid in the main text.

The view ordering, $S$, can be determined by minimizing a cost function. In this work, we have chosen to maximize the distance between points using electrostatic potential energies along the rows and columns. The criterion can be expressed as follows:

$$\Phi_{PR} = \sum_{j=1}^{n} \left( \underbrace{\sum_{i=1}^{m-1} \sum_{k=i+1}^{m} S(P_{ij}, P_{kj})}_{\phi_R(j)} \right) + \sum_{q=1}^{m} \left( \underbrace{\sum_{r=1}^{n-1} \sum_{s=r+1}^{n} S(P_{qr}, P_{qs})}_{\phi_C(q)} \right) \qquad [A1]$$

The function $\phi_R(j)$ denotes the electrostatic potential energy of all the points in different rows but in the same $j$-th column. Similarly, $\phi_C(q)$ denotes the electrostatic potential energy of all the points in different columns but in the same $q$-th row. Since the problem of optimal view-ordering can be solved using rectangular grid rather than jagged grid, we do not have to deal with point sets with different sizes among the rows or among the columns. Therefore, the cost function is simpler than the original cost function proposed in the main text. More importantly, due to the goals of making the signal modulation as incoherent as possible *in each echo train and among the same echoes across the trains* and *in different echo trains and in different echoes across the trains*, we have to have the two terms added as in Eq.[A1] rather than in a form of a ratio as in the original cost function Eq.[2].

Finally, the following modification should be introduced to the search algorithm in the main text:

Preliminary step: similar to the proposed algorithm in the main text.

Step 1: Find the index $j*$ for which $\phi_R(j)$ is maximum.

Step 2: Find the index $q*$ for which $\phi_C(q)$ is maximum.

Step 3: Look for lower energy configuration systematically by swapping $P_{ij}$ and $P_{q*j*}$ for all $i \neq q*$ and $j \neq j*$. If the rearranged configuration leads to lower $\Phi_{PR}$, exit and set the "flag" to 1. If none of the rearrangements leads to lower, we suggest using a point within $q*$-th row and a point within $j*$-th column and swap these points with other points in the grid. Specifically, we look for the point, $P_{q*r}$, that *maximizes* the following inner sum:

$$\sum_{\substack{s=1 \\ s \neq r}}^{m} S(P_{q*r}, P_{q*s}).$$

Similarly, we may also look for a point, $P_{ij*}$ that *maximizes* the following inner sum:

$$\sum_{\substack{k=1 \\ k \neq i}}^{n} S(P_{ij*}, P_{kj*}).$$

Then, we can successively swap each point of the two points, $P_{q*r}$ and $P_{ij*}$, with all other points in the grid to search for a better configuration. If the rearranged configuration leads to lower $\Phi$, exit and set the "flag" to 1. If none of the rearrangements leads to lower $\Phi$, then exit and set the "flag" to -1.

Steps 4 and 5: similar to the proposed algorithm shown in the main text.

### Example 1

It should be clear that the cost function and the search strategy presented in this appendix and in the main text can be immediately adapted to non-antipodally symmetric point set. As an illustration, we will show a visually convincing results of finding optimum configuration of a rectangular grid of 8 by 10. In this case, there are $4.8 \times 10^{107}$ distinct configurations. The lowest energy value of $\Phi_{PR}$ with non-antipodal electrostatic energy function, i.e., each occurrence of the function $S(P_v, P_w)$ in $\Phi_{PR}$ is replaced by $1/\|P_v - P_w\|$, was found to be 468.331889. The best configuration we obtained is shown in Figure 7 in two different views.

### Example 2

Ideally, we want the points in each column (or row) to be as uniform as possible but the electrostatic energy in each column or row will likely be higher than the gold standard (or the analytically exact spiral [66]) point set of the same size. By taking the ratio of the electrostatic energy of each column (or row) to the electrostatic energy of the analytically exact spiral point set of the same size and plotting the ratios of all the columns (or rows), we can easily inspect the variability or lack thereof, which indirectly related to the uniformity, of the columns (or rows). This method of evaluation and comparison will be used in this example.

The number of points (on the sphere) as deployed in a typical 3D radial MR experiment is in the order of thousands. Here, we will use a grid of (n=128,m=100), which has 12800 points and its corresponding number of distinct configurations in $10^{46642}$. We will apply the method of evaluation mentioned above on our optimum point set (configuration) and on the point set obtained by the commonly used methods such as the pseudo-random bit-reversal method [61] and the Golden Mean method [62], see Appendix B for a brief description of the 2D bit-reversal method. Based on the results shown in Figure 8, it is encouraging to see that the electrostatic energies of the proposed point set are less variable (in row or in column) and lower than those of the 2D bit-reversal method and the Golden Mean method. We also provided a more geometric comparison regarding the variability of the Voronoi areas and of the Voronoi circumferences of our point set and the Golden Mean method, which is known to be less uniform in distribution and hence higher variability in the above geometric measures, see Figure 9. Please refer to Appendix C in which we showed how we computed the average area and circumference of the spherical cap to get a rough approximation of the lower bound of the Voronoi circumference and the average value of the Voronoi area.

**APPENDIX B: 2D bit-reversal method**

For convenience, we include here a description of the 2D bit-reversal method. We assume the procedure of bit-reversal of an array of numbers is understood. Given a grid of m by n and an array of ordered numbers from 0 to m-1. First, bit-reverse this array and place it in the first column. Subsequently, the value of each component of the j-th column is equal to the corresponding component of the first column with ((j-1)*m) added to it. Second, bit-reverse the array of numbers in each row then, for each k-th row, cyclically shift the elements by k positions from right to left. The final grid is the desired solution of the 2D bit-reversal method.

**APPENDIX C: Area and circumference of a circular cap on the sphere.**

In this appendix, we present a simple expression of the circumference of a circular cone as a function of its area, which was used in Figure 9. Our derivation is built upon our previous work, i.e., page 839 of Ref. [15]. The unnormalized areal and circumferential measures for the case of the circular cone are given respectively by:

$$\Gamma = 2\pi \left( 1 - 1/\sqrt{1+r^2} \right),$$

and

$$\Lambda = 2\pi \left( r/\sqrt{1+r^2} \right),$$

where $r$ is the radius of the circular cone lying on the plane that is perpendicular to the unit sphere and the center of the cone is the point of contact between the plane and the unit sphere. After some manipulation, it can be shown that:

$$\Lambda = \sqrt{\Gamma(4\pi - \Gamma)}.$$

For example, if we assume that $\Gamma = 4\pi/12800$ then $\Lambda = 0.111068$. From the isoperimetric problems in calculus of variations, we know that a circle is the closed plane curve of a given length that encloses the largest area, and a circle is also the closed plane curve of the minimum length that encloses a given area. In fact, similar extension can be made about the surface area and the volume of the sphere, see e.g., page 224 of Ref.[67]. Therefore, the circumference, $\Lambda$, represents the minimum circumference and can serve as a lower bound for evaluating the quality of the Voronoi regions.

**ACKNOWLEDGMENTS**

C.G.K. dedicates this work to Pauline Toh and Eng Khoon Leong. Software related to this work will be made available through the following URL: http://sites.google.com/site/hispeedpackets. This work was supported in part by the National Institutes of Health IRCMH090912-01. E.Ö. was supported by the Department of Defense in the Center for Neuroscience and Regenerative Medicine (CNRM) and the Henry M. Jackson Foundation (HJF).

TABLE & FIGURE CAPTIONS

Table 1. A collection of 18 acquisition designs and the corresponding matrix condition numbers. The design matrices were constructed from the three-dimensional basis functions with u = 0.00827, which in turn depends on Δ and D. Here the diffusivity is chosen to be close to free diffusion of water in the brain. For example, Design #8 has (9,18,27,27) points in the (1st, 2nd, 3rd, 4th) shells respectively. This design has matrix condition number of 38.8 and A-optimal measure of $2.0 \times 10^7$. Further, the q-value at the first shell is 25.2 mm$^{-1}$. Design #18 is the square acquisition design.

| q-value (mm$^{-1}$) / Design \ Shell | 25.2 — 1 | 50.4 — 2 | 75.6 — 3 | 100.9 — 4 | 126.1 — 5 | 151.3 — 6 | 176.6 — 7 | 201.8 — 8 | 227.0 — 9 | Condition Number | A-optimal measure |
|---|---|---|---|---|---|---|---|---|---|---|---|
| 1 | 9 | 36 | 36 | | | | | | | 34.9 | $1.5 \times 10^7$ |
| 2 | 18 | 27 | 36 | | | | | | | 45.2 | $1.5 \times 10^7$ |
| 3 | 9 | 9 | 27 | 36 | | | | | | 39.8 | $3.9 \times 10^7$ |
| 4 | 9 | 18 | 18 | 36 | | | | | | 46.5 | $2.9 \times 10^7$ |
| 5 | 9 | 9 | 9 | 18 | 36 | | | | | 71.7 | $1.1 \times 10^8$ |
| 6 | 9 | 9 | 9 | 9 | 9 | 36 | | | | 68.1 | $1.2 \times 10^8$ |
| 7 | 27 | 27 | 27 | | | | | | | 63.0 | $2.0 \times 10^7$ |
| 8 | 9 | 18 | 27 | 27 | | | | | | 38.8 | $2.0 \times 10^7$ |
| 9 | 9 | 9 | 9 | 27 | 27 | | | | | 70.4 | $1.0 \times 10^8$ |
| 10 | 18 | 18 | 18 | 27 | | | | | | 64.3 | $3.0 \times 10^7$ |
| 11 | 9 | 9 | 18 | 18 | 27 | | | | | 47.3 | $5.7 \times 10^7$ |
| 12 | 9 | 9 | 9 | 9 | 18 | 27 | | | | 68.9 | $1.0 \times 10^8$ |
| 13 | 9 | 9 | 9 | 9 | 9 | 9 | 27 | | | 68.5 | $1.3 \times 10^8$ |
| 14 | 9 | 18 | 18 | 18 | 18 | | | | | 47.1 | $3.0 \times 10^7$ |
| 15 | 9 | 9 | 9 | 18 | 18 | 18 | | | | 69.5 | $1.1 \times 10^8$ |
| 16 | 9 | 9 | 9 | 9 | 9 | 18 | 18 | | | 70.2 | $1.1 \times 10^8$ |
| 17 | 9 | 9 | 9 | 9 | 9 | 9 | 9 | 18 | | 73.1 | $1.3 \times 10^8$ |
| 18 | 9 | 9 | 9 | 9 | 9 | 9 | 9 | 9 | 9 | 86.6 | $1.6 \times 10^8$ |

\* Δ = 16.32 ms, D = 2.1 x 10$^{-3}$ mm$^2$/s, u = 0.00827.

Table 2. The same collection of 18 acquisition designs as in Table 1 but the design matrices were now constructed from the basis functions with u = 0.000827. Here the value of the diffusivity was chosen to be low similar to the case of hindered diffusion.

| q-value (mm⁻¹) Design / Shell | 25.2 | 50.4 | 75.6 | 100.9 | 126.1 | 151.3 | 176.6 | 201.8 | 227.0 | Condition Number | A-optimal measure |
|---|---|---|---|---|---|---|---|---|---|---|---|
| | 1 | 2 | 3 | 4 | 5 | 6 | 7 | 8 | 9 | | |
| 1 | 9 | 36 | 36 | | | | | | | $7.6 \times 10^3$ | $2.0 \times 10^{13}$ |
| 2 | 18 | 27 | 36 | | | | | | | $7.1 \times 10^3$ | $2.1 \times 10^{13}$ |
| 3 | 9 | 9 | 27 | 36 | | | | | | $1.8 \times 10^3$ | $2.0 \times 10^{12}$ |
| 4 | 9 | 18 | 18 | 36 | | | | | | $2.0 \times 10^3$ | $2.2 \times 10^{12}$ |
| 5 | 9 | 9 | 9 | 18 | 36 | | | | | 689 | $4.0 \times 10^{11}$ |
| 6 | 9 | 9 | 9 | 9 | 9 | 36 | | | | 319 | $1.1 \times 10^{11}$ |
| 7 | 27 | 27 | 27 | | | | | | | $7.0 \times 10^3$ | $2.1 \times 10^{13}$ |
| 8 | 9 | 18 | 27 | 27 | | | | | | $1.9 \times 10^3$ | $2.0 \times 10^{12}$ |
| 9 | 9 | 9 | 9 | 27 | 27 | | | | | 664 | $3.8 \times 10^{11}$ |
| 10 | 18 | 18 | 18 | 27 | | | | | | $2.0 \times 10^3$ | $2.2 \times 10^{12}$ |
| 11 | 9 | 9 | 18 | 18 | 27 | | | | | 692 | $3.7 \times 10^{11}$ |
| 12 | 9 | 9 | 9 | 9 | 18 | 27 | | | | 308 | $1.0 \times 10^{11}$ |
| 13 | 9 | 9 | 9 | 9 | 9 | 9 | 27 | | | 159 | $3.1 \times 10^{10}$ |
| 14 | 9 | 18 | 18 | 18 | 18 | | | | | 761 | $4.0 \times 10^{11}$ |
| 15 | 9 | 9 | 9 | 18 | 18 | 18 | | | | 310 | $1.0 \times 10^{11}$ |
| 16 | 9 | 9 | 9 | 9 | 9 | 18 | 18 | | | 165 | $3.4 \times 10^{10}$ |
| 17 | 9 | 9 | 9 | 9 | 9 | 9 | 9 | 18 | | 93.0 | $1.2 \times 10^{10}$ |
| 18 | 9 | 9 | 9 | 9 | 9 | 9 | 9 | 9 | 9 | 80.3 | $8.8 \times 10^9$ |

\* $\triangle$ = 16.32 ms, D = 2.1 x 10⁻⁵ mm²/s, u = 0.000827.

| SHELL / INDEX | 1 | 2 | $\cdots$ | K |
|---|---|---|---|---|
| 1 | $P_{1\,1}$ | $P_{1\,2}$ | $\cdots$ | $P_{1\,K}$ |
| 2 | $P_{2\,1}$ | $P_{1\,2}$ | $\cdots$ | $P_{2\,K}$ |
| $\vdots$ | $\vdots$ | $\vdots$ | $\ddots$ | $\vdots$ |
| $n_K$ | $P_{n_K 1}$ | $P_{n_K 2}$ | $\cdots$ | $P_{n_K K}$ |
| $\vdots$ | $\vdots$ | $\vdots$ | $\vdots$ | |
| $\vdots$ | $\vdots$ | $\vdots$ | $\vdots$ | |
| $\vdots$ | $\vdots$ | $\vdots$ | | |
| $n_2$ | $P_{n_2 1}$ | $P_{n_2 2}$ | | |
| $n_1$ | $P_{n_1 1}$ | | | |

Figure 1. A jagged grid is used as a graphical representation to manage multiple-shell design; each column of data points is a collection of points on a spherical shell. Points in each row may be thought of as collections points around a radial line.

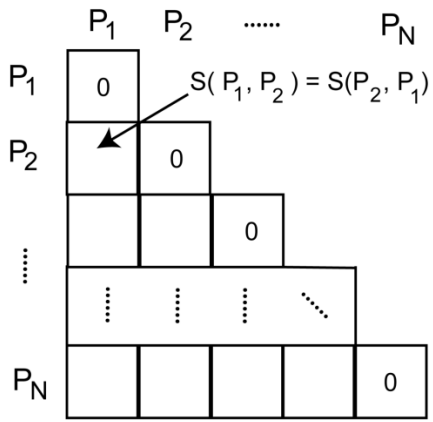

Figure 2. Computation of the cost function is greatly simplified by the use of the metric function S between two 'real' points, which is specifically designed to deal with point set that is endowed with antipodal symmetry. The lower triangular matrix shown above is used to keep the values of S.

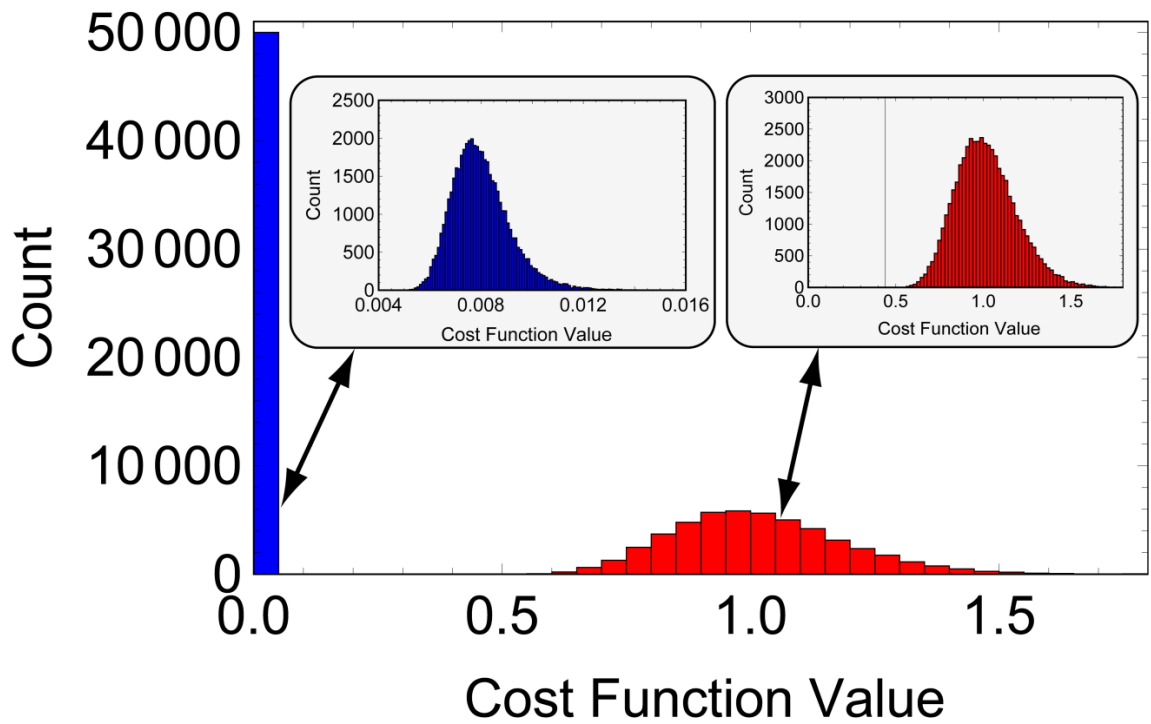

Figure 3. 50000 random permutations were generated to fill the 12 by12 grid. The initial cost function values of these 50000 samples are shown in the histogram that is color-coded in red. The histogram of the final cost function values of these 50000 samples is shown in blue. Inset shows the magnified version of these two histograms.

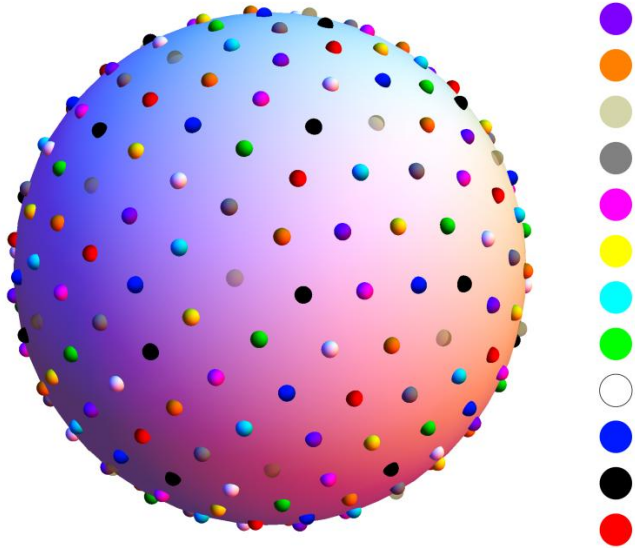

Figure 4. The points are generated from the example of a 12 by 12 grid. Every point in the same shell has the same color and each shell is assigned a distinct color and these colors are shown on the right. It can be seen that each set of points with the same color is nearly uniformly distributed on the sphere, which is related to the goal of criterion #1.

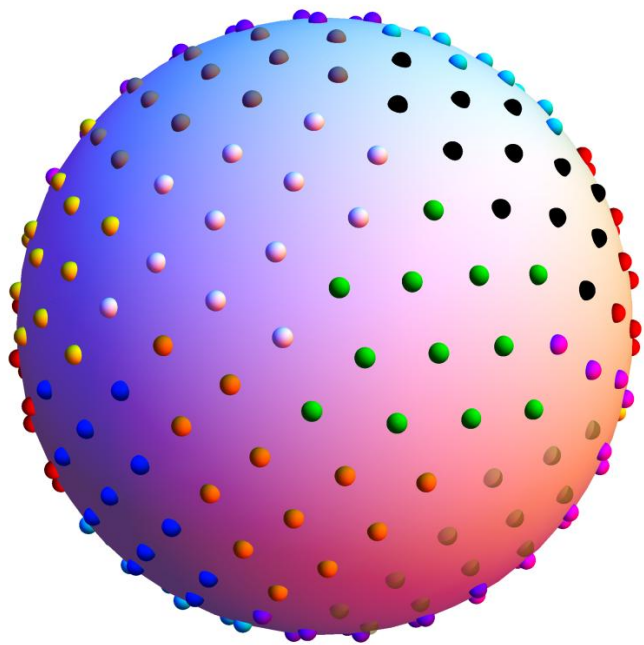

Figure 5. The clusters seen here are generated from the example of a 12 by 12 grid. Each row contains points with the same color and these points are designed to be close together in a form of a cluster so that when the points are projected to different shells we would have fulfilled the criterion #2, which is to provide the maximum coverage around each radial line. The problem of the boundary effect in which there might be two neighboring points with distinct colors but are moved to some common shell will not be an issue here because of criterion #1.

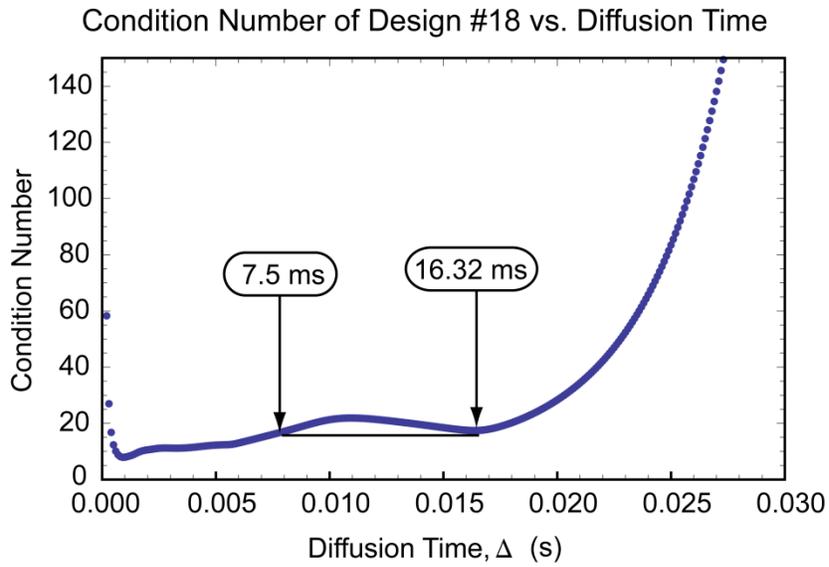

Figure 6. The condition number of the design matrix of Design #18 as a function of the diffusion time.

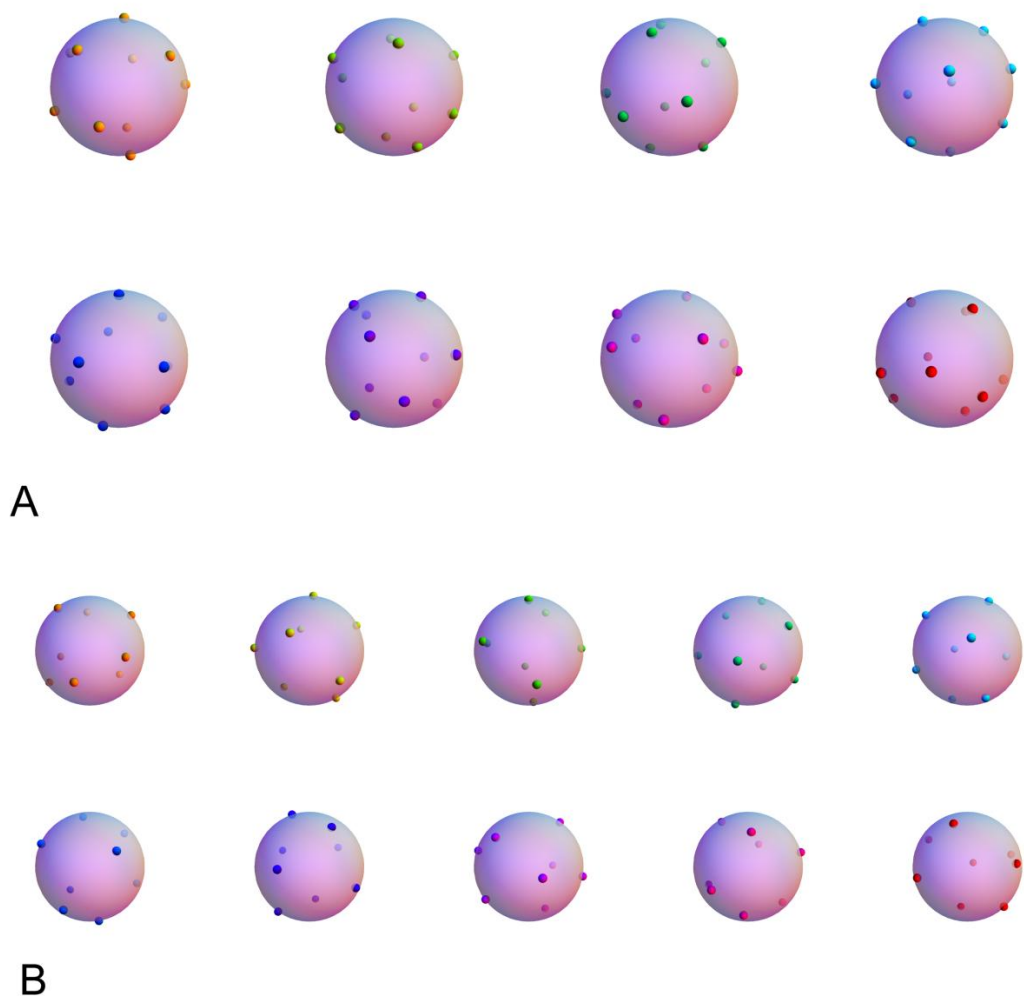

A

B

Figure 7. (A) There are 8 rows and each row of 10 points is shown here as points with the same hue on the translucent unit sphere. Points in different rows have different hues. (B) There are 10 columns and each column has 8 points. Each column of 8 points is shown here as points with the same hue on the translucent unit sphere. Again, points in different columns have different hues.

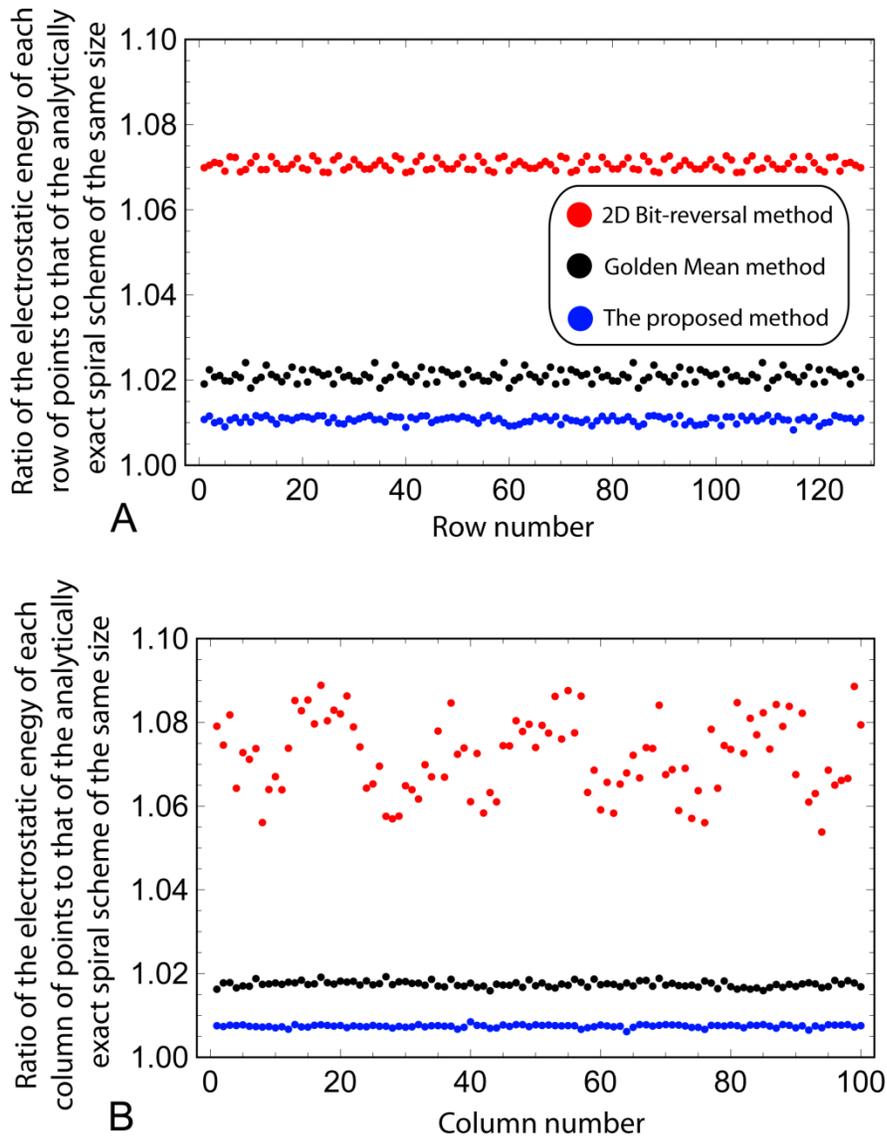

Figure 8. (A) The ratios of the electrostatic energy of points in each row of the 2D bit-reversal method (in red) and of the proposed method (in blue) to that of the point set of the same size (100 points) generated from the analytically exact spiral scheme. (B) The ratios of the electrostatic energy of points in each column of the 2D bit-reversal method (in red) and of the proposed method (in blue) to that of the point set of the same size (128 points) generated from the analytically exact spiral scheme.

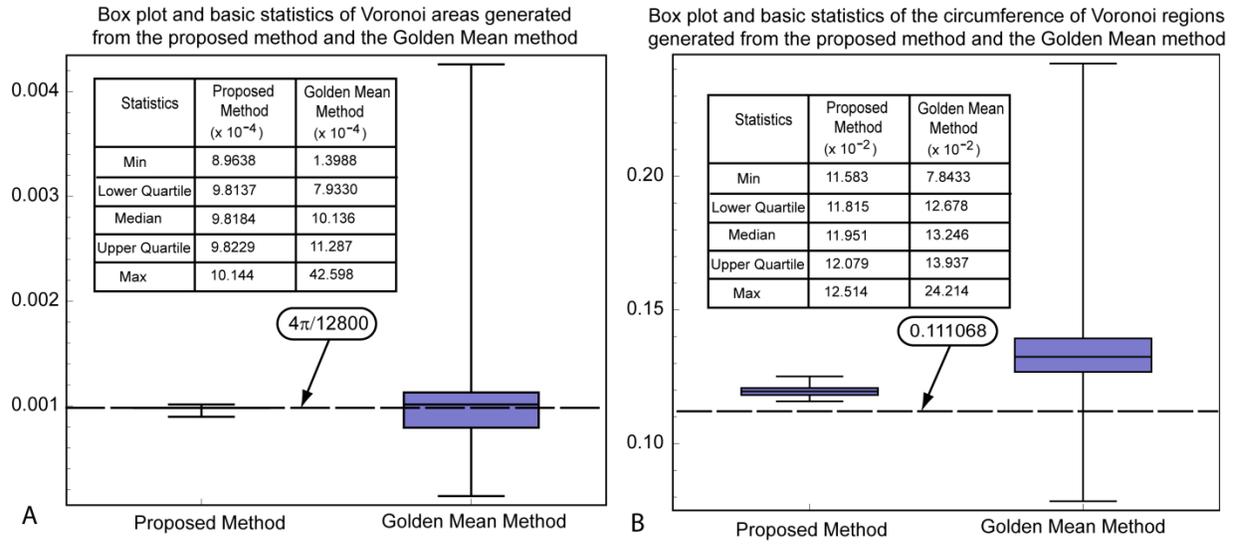

Figure 9. Box plots and basic statistics on the Voronoi areas (A) and circumferences (B) generated from the propose method, (analytically exact spiral scheme), and the Golden Mean method.